\documentclass[useAMS]{mn2e}
\usepackage{epsfig}
\DeclareMathSymbol{\gtrsim} {\mathrel}{AMSa}{"26}
\DeclareMathSymbol{\lessim} {\mathrel}{AMSa}{"2E}
\newcommand{\bd}{\begin{displaymath}} 
\newcommand{\ed}{\end{displaymath}} 
\newcommand{\kt}{k_{\rm B}T}
\newcommand{\be}{\begin{equation}}
\newcommand{\ee}{\end{equation}}
\newcommand{\sr}{\sigma_{\rm R}}
\newcommand{\rmd}{\,{\rm d}}
\newcommand{\kr}{\kappa_{\rm R}}
\newcommand{\gr}{g_{\rm rad}}
\title[OP data on CD for mean opacities and radiative accelerations]
{OP data on CD for mean opacities and radiative accelerations}
\author[M.J. Seaton]
{M.J. Seaton\\
Department of Physics and Astronomy, University
College London, London WC1E 6BT}
\begin{document}
\date{Accepted XXX. Received XXX; in original form XXX}

\pagerange{\pageref{firstpage}--\pageref{lastpage}} \pubyear{2004}

\maketitle

\label{firstpage}

\begin{abstract}
All monochromatic opacity data from the Opacity Project (OP),
together with all codes required for the calculation of mean
opacities and radiative accelerations for any required
chemical mixture, temperature and mass-density, are being put
on a 700 Mb CD which will be made generally available. The present
paper gives a concise summary of the contents of the CD. 
More complete documentation will be provided on the CD itself.
\end{abstract}
\section{Introduction}
Methods used by the Opacity Project (OP) for the calculation of mean 
opacities (Planck and Rosseland) were described in Paper I (Seaton {\em et al.}, 1994); further
work on the inclusion of
contributions from inner-shell transitions were discussed in Paper II
(Badnell and Seaton, 2003) for the six-element mix of Iglesias and Rogers 
(1995); comparisons with
OPAL (see Rogers \& Iglesias, 1992) for that mix in Paper III (Seaton and Badnell, 2004);
and up-dated results for all elements, with inclusion of inner-shell
contributions, in Paper IV (Badnell {\em et al}, 2004). The OP approach
to the calculation 
of radiative accelerations is discussed by Seaton (1997).

The basic monochromatic opacity data for the 17 elements
H, He, C, N, O, Ne, Na, Mg, Al, Si, S, Ar, Ca, Cr, Mn, Fe and 
Ni occupy 665 Mb as {\tt gzip} formatted files, or 868 Mb as binary files. 
Some improvements have been
made in the frequency-mesh which is used (see section 2.2.1 below)
which lead to improved accuracies in the numerical integrations
required to obtain mean opacities and radiative
accelerations. It is found that the integrations for Rosseland-means
never give numerical errors larger than 0.1\%. Improvements have been
made in the codes used to obtain mean opacities and accelerations
for any required chemical mixture, interpolated to any 
required values of temperature and mass-density. The interpolated
Rosseland mean opacities are shown to be numerically
correct to better than 1\% for all practical purposes.  

It is our intention to make all OP data generally available, both
on a 700 Mb CD which will be described in the present paper
and on a web OPserver which will be described by Mendoza {\em et al.}
(2004).

The CD contains all data required for the calculation of OP mean opacities
and radiative accelerations, together
with full documentation and all required codes, and is available
on request\footnote{Requests to: Claude Zeippen 
(\mbox{Claude.Zeippen@obspm.fr}); or Anil Pradhan(
\mbox{pradhan.1@osu.edu})}. It contains a {\tt tar} file of compressed ({\tt gzip})
files. The file {\tt OPCD.ps.gz} provides full documentation,
including information about
processing of all other data on the {\tt tar} file. 

The user is urged to read the file {\tt OPCD.ps}  {\em before}  
decompressing any other files.

The present paper gives a concise summary of the contents of the CD.
Full documentation is given in the file {\tt OPCD.ps}.
The CD includes input and output files for some test runs.
\section{Basic definitions}
\subsection{Monochromatic opacities}
The monochromatic opacity cross-section for element $k$ is $\sigma_k(u)$ where
$u=h\nu/(\kt)$ and $k_{\rm B}$ is Boltzmann's constant. The 
correction-factor for stimulated emission, $[1-\exp(-u)]$, is
{\em not} included\footnote{We here
consider the quantity actually tabulated in the OP archives. A
number of previous papers have defined $\sigma_k$ {\em with} inclusion
of the stimulated emission factor.} 
in the definition of $\sigma_k$. For a mixture
of elements let element $k$ have abundance by number-fraction of
$f_k$, $\sum_k f_k=1$. The cross-section for the mixture is
\be \sigma(u)=\sum_k f_k \sigma_k(u). \ee
\subsection{The Rosseland mean}
The Rosseland-mean cross-section is
\be \frac{1}{\sr}=\int_0^\infty \frac{F(u)}{\sigma(u)[1-\exp(-u)]}
\rmd u \ee
where
\be F(u)=[15/(4\pi^4)]u^4\exp(-u)/[1-\exp(-u)]^2. \ee
The Rosseland-mean opacity, per unit mass, is $\kr=\sr/\mu$ where 
$\mu$ is mean atomic weight.

Numerical values for cross-sections $\sigma$ will be given in 
atomic units ($a_0^2$) and values for opacities $\kr$ in cgs units
(cm$^2$ g$^{-1}$).
\subsubsection{The frequency variable}
It is convenient to use the frequency-variable\footnote{Some previous 
papers have taken $\sigma$ to be
defined {\em with} inclusion of the stimulated emission factor, and
$v$ to be defined by $v=\int F(u) \rmd u $.}
\be v(u)=\int_0^u \frac{F(u)}{[1-\exp(-u)]} \rmd u \ee
giving
\be \frac{1}{\sr}=\int_0^{v_\infty} \frac{1}{\sigma(u)} \rmd v \ee
where $v_{\infty}=v(u\rightarrow \infty)$ 
\subsection{Radiative accelerations}
The radiative acceleration for element $k$ in a stellar interior is
\be  \gr=(1/c)(\mu/\mu_k){\cal F}\kr\gamma_k \ee
where $c$ is the speed of light, $\mu_k$ the atomic mass of element
$k$ and
\be {\cal F}=\pi B(T_{\rm eff})(R_\star/r)^2 \ee
where $T_{\rm eff}$ is the effective temperature,
\be B(T)=2(\pi\kt)^4/(15c^2h^3), \ee
$R_\star$ is the radius of the star and $r$ the distance from
the centre of the star. In (6),
\be \gamma_k=\int \frac{\sigma^{\rm mta}_k}{\sigma} \rmd v \ee
where $\sigma^{\rm mta}_k$ is a cross-section for momentum-transfer
to the atom,
\be \sigma^{\rm mta}_k=\sigma_k(u)[1-\exp(-u)]-a_k(u). \ee
In (10), $a_k(u)$ is a correction factor to take out contributions
from scattering processes and momentum transfer to electrons
in photo-ionization processes.

Values of $\gr$ are given in cgs units (cm s$^{-2}$). The quantity
$\gamma_k$ is dimensionless.

One is usually interested in the case in which the abundance of the
selected element $k$ varies with depth in a star due to the
operation of diffusion processes. The codes allow for the
possibility of varying the abundance of $k$ by a factor $\chi$, keeping
the relative abundances of all other elements unchanged. 
\section{Archived data files}
All data are tabulated on a mesh of values of temperature $T$ and
electron-density $N_e$. The indices {\tt ite} and {\tt jne} are
defined by
\be \mbox{\tt ite}=40\times \log(T),\,\,\, \mbox{jne}=4\times \log(N_e).\ee
All data on the CD are with the ``m'' mesh, even values of {\tt ite}
and {\tt jne}, $\Delta${\tt ite}=$\Delta${\tt jne}$=2$.

The cross-sections are tabulated for {\tt ntot}$=10000$ 
equally-spaced values of $v$. The $\sigma_k$ data are in files
{\tt mzz.ttt} and the $a_k$ are in files {\tt azz.ttt} where {\tt zz} 
is a two-digit
number specifying nuclear charge ({\tt zz}$=01$ for H) and
{\tt ttt} is a three-digit number specifying {\tt ite}. Data for all
{\tt jne} included are contained in each files {\tt mzz.ttt} or {\tt azz.ttt}.
Files {\tt mzz.mesh} give values of $u$ for the tabulated values
of $v$. 

The OP codes use unformatted (binary) data files, for economy
in storage and speed of input/output. However, the CD contains
files which are formatted and compressed ({\tt gzip}) for ease of
transfer between machines. The procedures to be used
for converting to binary are described in the file {\tt OPCD.ps}.
 
Some summary data, described in the documentation, are given in
files {\tt mzz.index}, {\tt mzz.ion}, {\tt mzz.smry}.
\section{Codes for Rosseland-means}
\subsection{The code {\tt mixv.f}} The code {\tt mixv.f} reads a 
list of abundances $f_k$
for a mixture, obtains the cross-sections $\sigma$ and produces an output
file with values of mass-density $\rho$ and Rosseland means $\kr$
for all $(T,N_e)$ mesh points.
\subsection{The code {\tt opfit.f}} The code {\tt opfit.f} reads an output file from {\tt mixv.f}
and a list of values of $(\log(T),\log(\rho))$ for a stellar model
and produces a  file giving values of 
\be \log(\kr),\,\,\, \partial\log(\kr)/\partial\log(T)
,\,\,\,\partial\log(\kr)/\partial\log(\rho). \ee
The code uses refined techniques of bi-cubic spline interpolations.
It has an optional facility (rarely required) to smooth the data
before doing the interpolations, and an optional feature to produce
output tables in OPAL format.
\subsection{The code {\tt mx.f}} 
The code {\tt mx.f} produces output similar to that from {\tt opfit.f}
in a one-step process, using less refined
bicubic interpolation techniques. It has a facility that abundances
can be changed at any point during a run for a particular stellar
model.
\subsection{The code {\tt mixz.f}}
The code {\tt mixz.f} produces monochromatic opacities for any
specified mixture of metals. The output files are stored as 
a "fictitious" metal, {\tt mzz.ttt} with {\tt zz} a two-digit
number greater than 28 (data for Nickel are stored in files
{\tt m28.ttt}).
\subsection{The code {\tt mxz.f}}
The code {\tt mxz.f} is similar to {\tt mx.f} but uses the
files for a metal-mixture from {\tt mixz.f}. It has a facility that,
at each depth point, the values of $X$ and $Z$ can be changed.
\subsection{Choice of mixture codes}
Use of {\tt mixv.f} plus {\tt opfit.f} can be convenient if the same
mixture is to be used for many stellar models. The use of splines
in {\tt opfit.f} ensures that the derivatives in (12) are continuous.
It is shown in Paper IV that, for all cases likely to be of practical
interest, the interpolations using {\tt opfit.f} do not introduce
any errors as large as 1\%.

Use of {\tt mx.f} is particularly convenient if there are
changes in composition depending on depth in a stellar model.
In occasional difficult cases use of {\tt mx.f} can give a lack
of smoothness in the derivatives.

The test runs of {\tt mixv.f, opfit.f} and {\tt mx.f} are for S92 
abundances (see Paper I), $\log(T)=6.0$ to 7.3
and $\log(R)=-1.75$, roughly corresponding to conditions in the solar 
radiative interior. Use of {\tt mx.f} gives values of $\log(\kr)$ and
$\partial\log(\kr)/\partial\log(T)$ which appear to be 
smooth and in agreement with values from {\tt opfit.f} but there
is some apparent lack of smoothness in $\partial\log(\kr)/\partial\log(\rho)$
from {\tt mx.f}, as shown on Figure 1.

The codes {\tt mixz.f} and {\tt mxz.f} were used in Paper IV for
the case a more realistic solar model in which,
below the base of the convection zone, $X$ and $Z$ varied continuously
as functions of depth.
\section{Codes for radiative accelerations}
\subsection{The code {\tt accv.f}}
The code {\tt accv.f} has a structure similar to that of {\tt mixv.f}.
It reads a list of abundances and specification of the selected
element $k$, reads data specifying a range of values of the
abundance-multiplier $\chi$,
reads the files {\tt mzz.ttt} and {\tt azz.ttt} and
produces an output file giving, for each $(T,N_e)$ mesh point,
values of
\be \kr,\,\,\,\partial\kr/\partial\chi,\,\,\,\gamma,\,\,\, 
\partial\gamma/\partial\chi \ee
for each value of $\chi$.
\subsection{The code {\tt accfit.f}}
The code {\tt accfit.f} is similar to {\tt opfit.f} using bi-cubic
splines. It reads a file output from {\tt accv.f}, a list of 
values of $\chi$, the effective temperature $T_{\rm eff}$ for
a stellar model and a list of values of $\log(T)$, $\log(\rho)$ and 
$r/R_\star$. It produces a file giving value of
\be \log(\kr),\,\,\,\log(\gamma),\,\,\,\log(\gr) \ee
as functions of $\chi$.
\subsection{The code {\tt ax.f}}
The code {\tt ax.f} has a structure similar to that of {\tt mx.f}.
Its use is convenient when the mixture of elements depends
on depth in a stellar model.
\subsection{Choice of codes for accelerations}
Use of {\tt accfit.f} provides the most accurate interpolations. 
The code {\tt ax.f} can be useful for cases in which abundances
vary with depth. 

The test case is for Argon in a stellar model with $T_{\rm eff}=10^4$K.
Figure 2 shows $\gr$ for $\chi=0.01,\,1.00$ and 100. The deep minimum
in the vicinity of $\log(T)=5.45$ occurs when the Argon is going
through a Neon-like ionization stage. It is seen
that $\gr$ from {\tt ax.f} shows some lack of smoothness for the
rather extreme case of $\chi=0.01$
\section*{References}
Badnell N.R., Seaton M.J., 2003. J. Phys. B, 36,4367 (Paper~II)\\
Badnell N.R., Bautista M.A., Butler K., Delahaye F., Mendoza C.,
Palmeri P., Zeippen C.J., Seaton M.J., 2004. MNRAS, to be submitted  
(Paper~IV)\\
Iglesias C.A., Rogers, F.J., 1995, ApJ 443, 469\\
Mendoza, C. {\em et al.}, 2004, MNRAS, to be submitted\\
Rogers, F.J., Iglesias, C.A., 1992. ApJS 79,507, 1992\\
Seaton, M.J., 1997, MNRAS, 289, 700\\
Seaton, M.J., Badnell, N.R., 2004, MNRAS, in press (Paper~III)\\
Seaton, M.J., Yu Yan, Mihalas D., Pradhan, A.K., 1994, MNRAS,
266, 805 (Paper~I)\\
\clearpage
\begin{center}
\epsfig{file=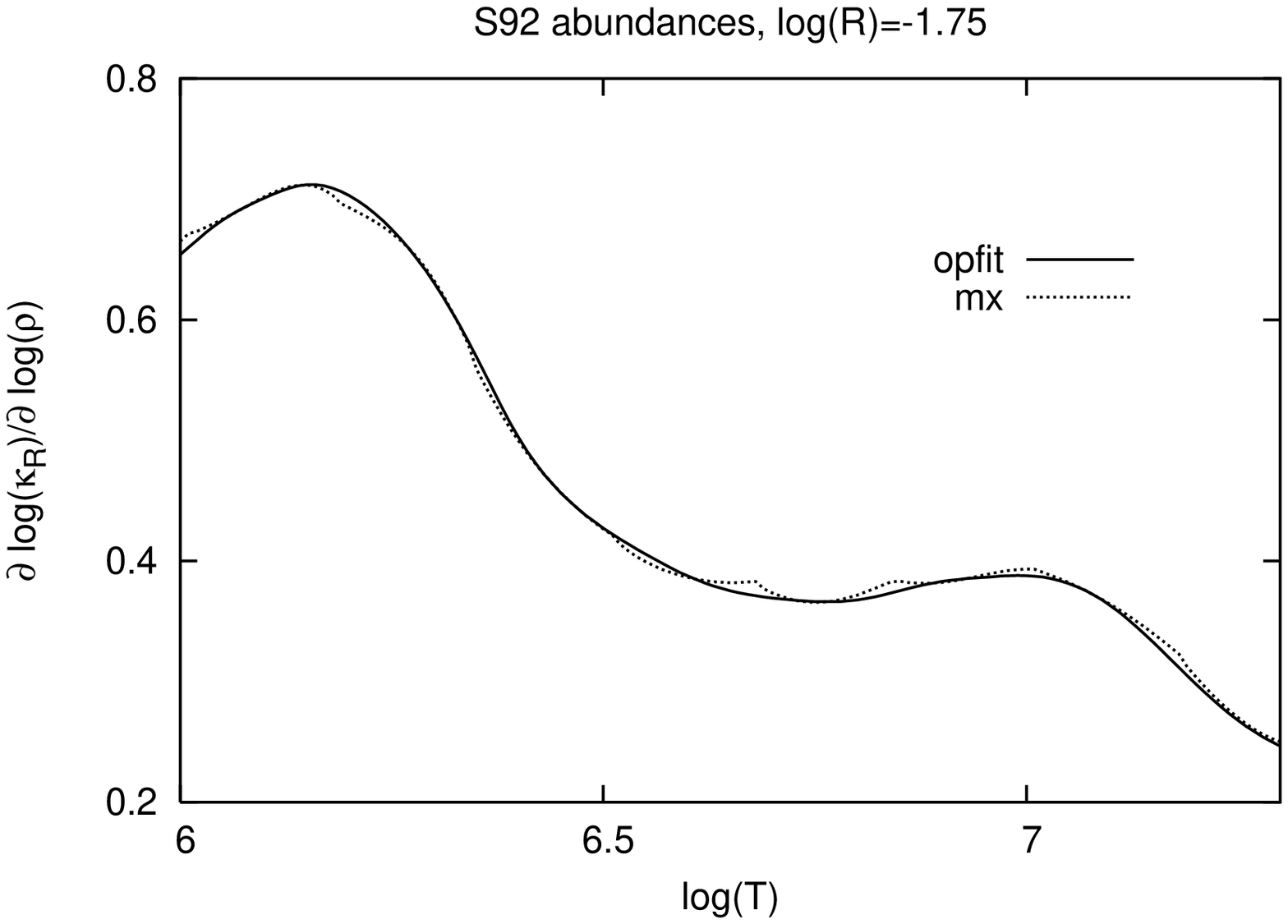,angle=270,width=10cm}
\end{center}
\vspace{5mm}
{\bf Figure 1} $\partial\log(\kr)/\partial\log(\rho)$ for a model
with S92 abundances and $\log(R)=-1.75$. Results obtained using
the codes {\tt opfit.f} and {\tt mx.f}.
\begin{center}
\epsfig{file=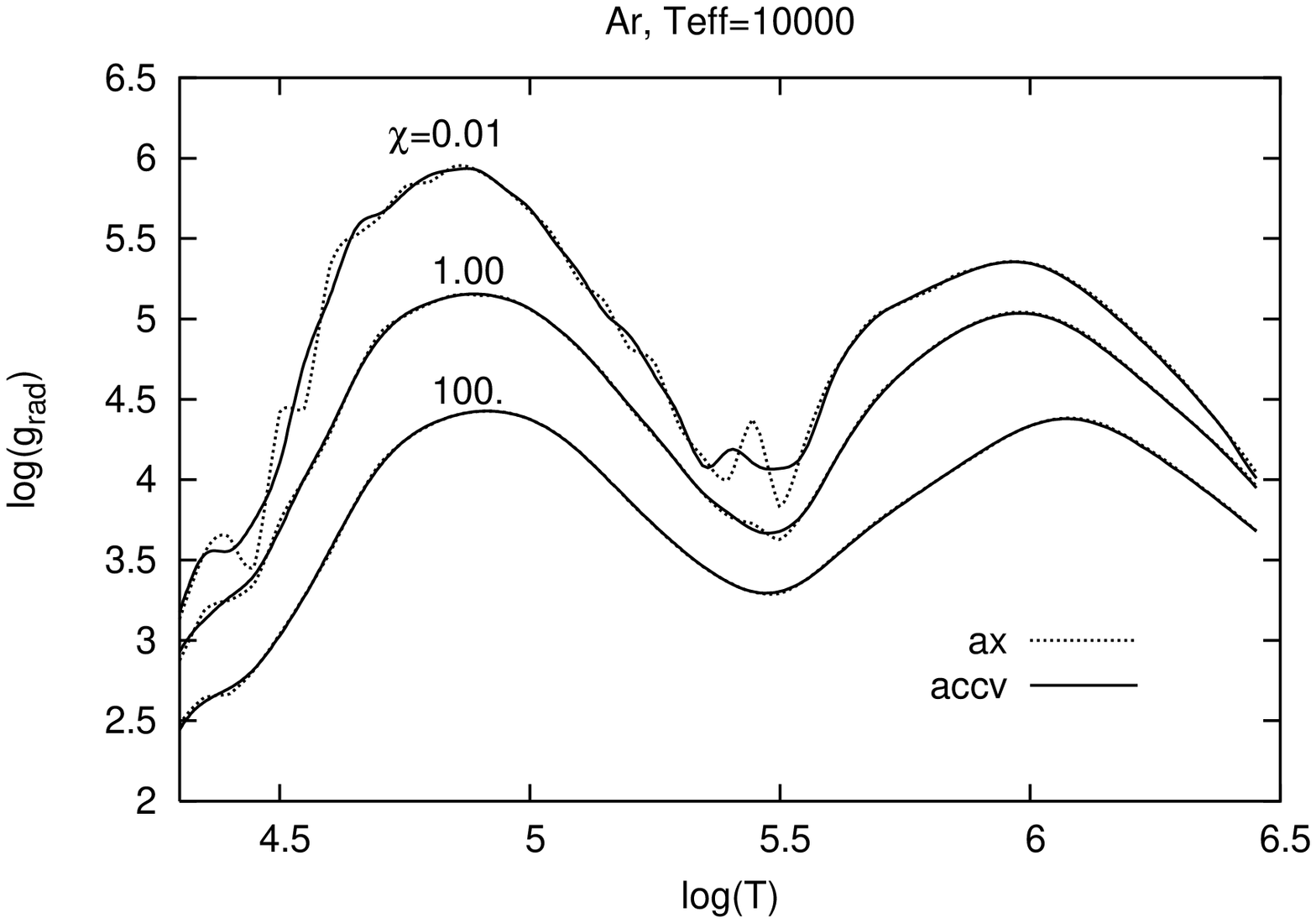,angle=270,width=10cm}
\end{center}
\vspace{5mm}
{\bf Figure 2} Radiative accelerations for Ar in a model with
$T_{\rm eff}=10^4$K. Results from codes {\tt accv.f} and
{\tt ax.f} with abundance multipliers $\chi=0.01,\,1.0$ and 100.
\label{lastpage}
\end{document}